\documentclass[letterpaper,12pt]{article}

\usepackage{authblk}
\usepackage{graphicx}
\usepackage{braket}
\usepackage{amsmath}
\usepackage{amssymb}
\usepackage{mathtools}
\usepackage{caption}
\usepackage{subcaption}
\usepackage{hyperref}
\usepackage{leftindex}
\usepackage[square,numbers,compress]{natbib}
\usepackage{color}

\usepackage[margin=1in,letterpaper]{geometry}
\usepackage{multirow}

\providecommand{\keywords}[1]
{
	\small	
	\textbf{\textit{Keywords---}} #1
}

\numberwithin{equation}{section}

\title{\vspace{-1.5cm}\textbf{\large Coherent Entanglement between Massive Oscillators from the Quantized Gravitational Field}}
\author[1]{\normalsize Anom Trenggana \footnote{Electronic address: gstanomagung@gmail.com}}
\author[1,2]{\normalsize Freddy P. Zen \footnote{Electronic address: fpzen@fi.itb.ac.id}}
\affil[1]{\textit{\normalsize Theoretical Physics Laboratory, THEPi (Theoretical High Energy Physics) Division, Faculty of Mathematics and Natural Sciences, Institut Teknologi Bandung, Bandung 40132, West Java, Indonesia}}
\affil[2]{\textit{\normalsize Indonesian Center for Theoretical and Mathematical Physics (ICTMP), Institut Teknologi Bandung, Bandung 40132, West Java, Indonesia}}
\date{(\today)}

\begin{document}
\maketitle
\begin{abstract}
In this work, we investigate how entanglement between two massive quantum particles trapped in harmonic oscillator potentials is generated after they interact with propagating modes of the quantized gravitational field. Specifically, we consider only the entanglement arising from the coherent contribution of the gravitational field to investigate its causal structure. We employ the Feynman–Vernon influence functional approach, within which the contribution of the coherent gravitational field to entanglement generation can be traced to the commutation relation of the gravitational field. Our analysis shows that entanglement does not arise instantaneously after the interaction begins. Instead, there is a time delay proportional to the separation between the two particles before entanglement begins to develop. This result shows that the entanglement arising from the coherent contribution of the gravitational field exhibits a causal structure. The magnitude of the resulting entanglement is generally very small. However, when the initial quantum states of the two particles are prepared as separable squeezed states, the resulting entanglement can be enhanced, with the magnitude of the enhancement depending exponentially on the squeezing parameter.
\end{abstract}


\newpage

\tableofcontents

\section{Introduction}\label{sec1}
The success of the \textit{Laser Interferometer Gravitational-Wave Observatory} (LIGO) detector in observing gravitational waves in 2016 \cite{LIGO} has raised hopes among scientists that, in the future, other phenomena and unique properties related to gravity may also be detected. One of the issues that has recently attracted considerable attention is whether the gravitational field possesses quantum properties. Answering this question is important because the detection of quantum properties or the quantized nature of the gravitational field could provide a foundation for the existence of a theory of quantum gravity.

Various efforts to provide experimental evidence for the quantum nature of gravity are still actively being pursued. Several interesting proposals regarding methods that can be used to detect the quantum properties of gravity have been put forward \cite{PWZ, Maity, Cho, Marshman, Bose, Sen, Bhatta, Penrose, Kanno1, Kanno2, Ikeda, Kanno3, Anom, Anom2, Nandi}. One particularly interesting proposal is the detection of the quantum nature of gravity through its role in generating entanglement between matter degrees of freedom. This approach is based on the idea that if gravitational interaction is non-classical, then it can act as a mediator capable of intrinsically correlating quantum systems. This idea is related to the fact that entanglement cannot be generated through \textit{local operations and classical communication} (LOCC) \cite{Bennet}. Therefore, a mediator that acts only as a classical channel cannot generate entanglement between quantum systems. The method of testing the quantum nature of gravity through this mechanism is commonly known as \textit{Quantum Gravity Induced Entanglement of Masses} (QGEM).

Several theoretical studies have shown that entanglement can arise as a consequence of the gravitational interaction between massive particles \cite{Bose2, Mzumdar, Bose3, Rufo, Toros, Schut, Mukherjee}. In these studies, entanglement arises from the interaction with the quantized gravitational field, which induces energy shifts in the massive particles and causes their quantum states to evolve into entangled states. In general, the gravitational field is treated as a fully quantized field, using the quantization scheme developed by S. N. Gupta \cite{Gupta}, while the relevant dynamics are dominated by the non-propagating sector. Consequently, the effects of the interaction are typically expressed in terms of an effective interaction, so that information about the propagation and dynamics of the gravitational field does not appear explicitly in the evolution of the system. Thus, in these approaches, the causal propagation of the gravitational field has not been explicitly incorporated into the analysis of entanglement generation.

On the other hand, several other studies have explicitly considered causal aspects of gravity-induced entanglement, namely \cite{Christodoulou, Bengyat}. These studies show that when propagating modes of the gravitational field are taken into account, while explicitly preserving the locality of the interaction, entanglement does not arise instantaneously but can only develop after a time delay. This indicates the presence of a causal structure in the generation of entanglement due to the interaction with the gravitational field. This causal structure was further investigated by Mitrakos \textit{et al.} \cite{Mitrakos}. They showed that the presence of a time delay in the generation of entanglement through gravity can provide support for the existence of gravitons, in contrast to entanglement generation without a time delay. Within the framework of the quantized gravitational field, this causal structure is associated with the commutator contribution of the gravitational field.

In the context of interactions between quantum systems, when one system is regarded as an environment for another system, correlations within the environment can influence the dynamics of the system interacting with it. In the framework of quantum fields, the commutation relation of the field operators plays an important role in determining the causal structure of these correlations. In the case of interactions with a quantized gravitational field, approaches that treat the gravitational field as an environment interacting with matter systems have been discussed in several studies \cite{Toros2, Cheng}. In these approaches, the commutation relation of the gravitational field plays a role in determining its influence on the dynamics of the matter system. However, these studies do not specifically investigate the generation of entanglement resulting from such interactions.

For this reason, in this work, we combine the treatment of the gravitational field as an environment for the matter systems with the propagating modes of the gravitational field to investigate the generation of entanglement and its causal structure. Using the Feynman–Vernon approach \cite{Breuer, Feynman, Feynman2}, the influence of the gravitational field on the matter systems can be analyzed through field correlations. In particular, this work focuses on the commutator contribution of the gravitational field to investigate the causal structure of entanglement generation. Within the Feynman–Vernon approach, the commutator contribution of the gravitational field is associated with the coherent contribution of the environment. Specifically, we investigate the generation of entanglement between two matter systems due to their interaction with an environment consisting of propagating modes of the gravitational field, as well as the causal structure arising from the commutator contribution of the gravitational field. It should be noted that the entanglement considered in this work specifically originates from the coherent contribution of the gravitational field and is not intended to represent the complete entanglement of the matter systems. The matter systems considered here consist of two massive particles, each trapped in a harmonic oscillator potential. The two particles interact with the quantized gravitational field through the coupling between the gravitational field and the matter energy-momentum tensor.

The structure of this paper is as follows. Section \ref{sec1} presents the introduction, including the background, objectives, and organization of the paper. Section \ref{sec2} discusses the quantization of the propagating gravitational field and the system model consisting of two massive particles trapped in harmonic oscillator potentials. Section \ref{sec3} then discusses the density matrix of the system after interacting with the propagating modes of the gravitational field, using the operator approach within the Feynman–Vernon influence functional framework. Based on the density matrix obtained in Section \ref{sec3}, Section \ref{sec4} analyzes the entanglement generated by the coherent contribution of the propagating modes of the gravitational field and evaluates the corresponding amount of entanglement. The final section presents the conclusions of this work. In addition, the paper includes two appendices containing details of the calculations used in this work.

\section{Graviton–Matter System: Two Quantum Harmonic Oscillators}\label{sec2}
In this section, we introduce a system model consisting of two massive oscillating particles interacting with a quantized gravitational field through its propagating modes. This section is divided into three parts. The first part discusses the quantization of the gravitational field, the second part introduces the system model, and the final part presents the form of the interaction Hamiltonian.

\subsection{Quantization of Linearized Gravitational Field}
In this subsection, linear gravitational perturbations propagating on a Minkowski spacetime background are considered. Accordingly, the spacetime metric can be written in the form
\begin{eqnarray}
ds^2=-dt^2+\big(\delta_{ij}+h_{ij}\big)\,dx^i\,dx^j.
\end{eqnarray}
Here, $t$ denotes the time coordinate, while $x^i$ represents the spatial coordinates, with $i=1,2,3$. The quantity $\delta_{ij}$ is the Kronecker delta, and $h_{ij}$ represents a small perturbation to the flat background metric. This perturbation satisfies the transverse-traceless (TT) conditions, namely $\partial_j h_{ij}=0$ and $h_{ii}=0$. We then consider the Einstein–Hilbert action up to second order in the perturbation $h_{ij}$, given by
\begin{eqnarray}\label{eq:2.2}
S^{(2)}_g&=&\frac{M^2_{pl}}{2}\,\,\int\,d^4 x\,\sqrt{-g}\,R\nonumber\\
&\simeq& \frac{M^2_{pl}}{2}\,\,\int\,d^4 x\,\Big[\dot{h}^{ij}\,\dot{h}_{ij}-h^{ij,k}\,h_{ij,k}\Big],
\end{eqnarray}
where $M_{pl}^2=1/(8\pi G)$. Here we employ natural units, $\hbar = c = 1$. The dot denotes a derivative with respect to time, while a comma followed by a spatial index denotes a spatial derivative.

To facilitate quantization, the gravitational perturbation field is expanded in Fourier space as
\begin{eqnarray}\label{eq:2.3}
h_{ij}(t,\boldsymbol{x})=\frac{2}{M_{pl}\sqrt{V}}\,\sum_{\boldsymbol{k},s}\,h^s_{\boldsymbol{k}}(t)\,e^{i\boldsymbol{k}\cdot\boldsymbol{x}}\,\epsilon^{s}_{ij}(\boldsymbol{k}),
\end{eqnarray}
Here, $\boldsymbol{k}$ denotes the wave vector, and $s$ denotes the polarization index. For linear polarization, this index takes two independent values, $s=+,\times$. Here, $V=L_xL_yL_z$ is the normalization volume of the system. By imposing periodic boundary conditions, the components of the wave vector become discretized as $\boldsymbol{k}=\left(\frac{2\pi n_x}{L_x}, \frac{2\pi n_y}{L_y}, \frac{2\pi n_z}{L_z}\right)$ where $n_x,n_y,n_z$ are integers. The polarization tensor $\epsilon^{s}_{ij}(\boldsymbol{k})$ satisfies
\begin{eqnarray}
\sum_{s}\,\,{\epsilon^{s}_{ij}}^*(\boldsymbol{k})\,\epsilon^{s}_{kl}(\boldsymbol{k})=\frac{1}{2}\,\left(P_{ik}P_{jl}+P_{il}P_{jk}-P_{ij}P_{kl}\right),
\end{eqnarray}
where $P_{ij}=\delta_{ij}-\bar{k}_i\bar{k}_j$ is the projection tensor and $\bar{k}_i=k_i/k$ denotes the component of the unit vector. Accordingly, the polarization tensor satisfies the orthonormality condition $\epsilon^{A*}_{ij}(\boldsymbol{k})\,\epsilon^{B}_{ij}(\boldsymbol{k})=\delta^{AB}$.

Substituting the Fourier expansion into the action in equation (\ref{eq:2.2}), the equation of motion for each gravitational mode can be obtained as,
\begin{eqnarray}\label{eq:2.4}
\ddot{h}^s_{\boldsymbol{k}}(t)+k^2\,h^s_{\boldsymbol{k}}(t)=0.
\end{eqnarray}
This equation shows that each Fourier mode of the gravitational perturbation is equivalent to a quantum harmonic oscillator with angular frequency $\omega_k=|\boldsymbol{k}|=k$. Therefore, the quantity $h^s_{\boldsymbol{k}}(t)$ can be promoted to a quantum operator and expressed in terms of creation and annihilation operators as
\begin{eqnarray}\label{eq:2.6}
\hat{h}^{s}_{\boldsymbol{k}}(t)=\hat{c}^s_{\boldsymbol{k}}\,\,u_k(t)+\hat{c}^{s\dagger}_{-\boldsymbol{k}}\,\,u^*_k(t).
\end{eqnarray}
This form ensures that the gravitational field remains real, i.e. $\hat{h}_{ij}(t,\boldsymbol{x})=\hat{h}_{ij}^\dagger(t,\boldsymbol{x})$. The annihilation operator $\hat{c}^s_{\boldsymbol{k}}$ and creation operator $\hat{c}^{s\dagger}_{\boldsymbol{k}}$ satisfy the canonical commutation relation $[\hat{c}^A_{\boldsymbol{k}}, \hat{c}^{B\dagger}_{\boldsymbol{p}}]=\delta^{AB}\delta_{\boldsymbol{k},\boldsymbol{p}}$, while $u_k(t)$ is the mode function solving equation (\ref{eq:2.4}), given by
\begin{eqnarray}
u_k(t)=\frac{1}{\sqrt{2\,k}}\,e^{-i\,k\,t}.
\end{eqnarray}

\subsection{Matter System: Two Quantum Harmonic Oscillators}
\begin{figure}[htbp]
   \centering
   \includegraphics[width=0.7\textwidth]{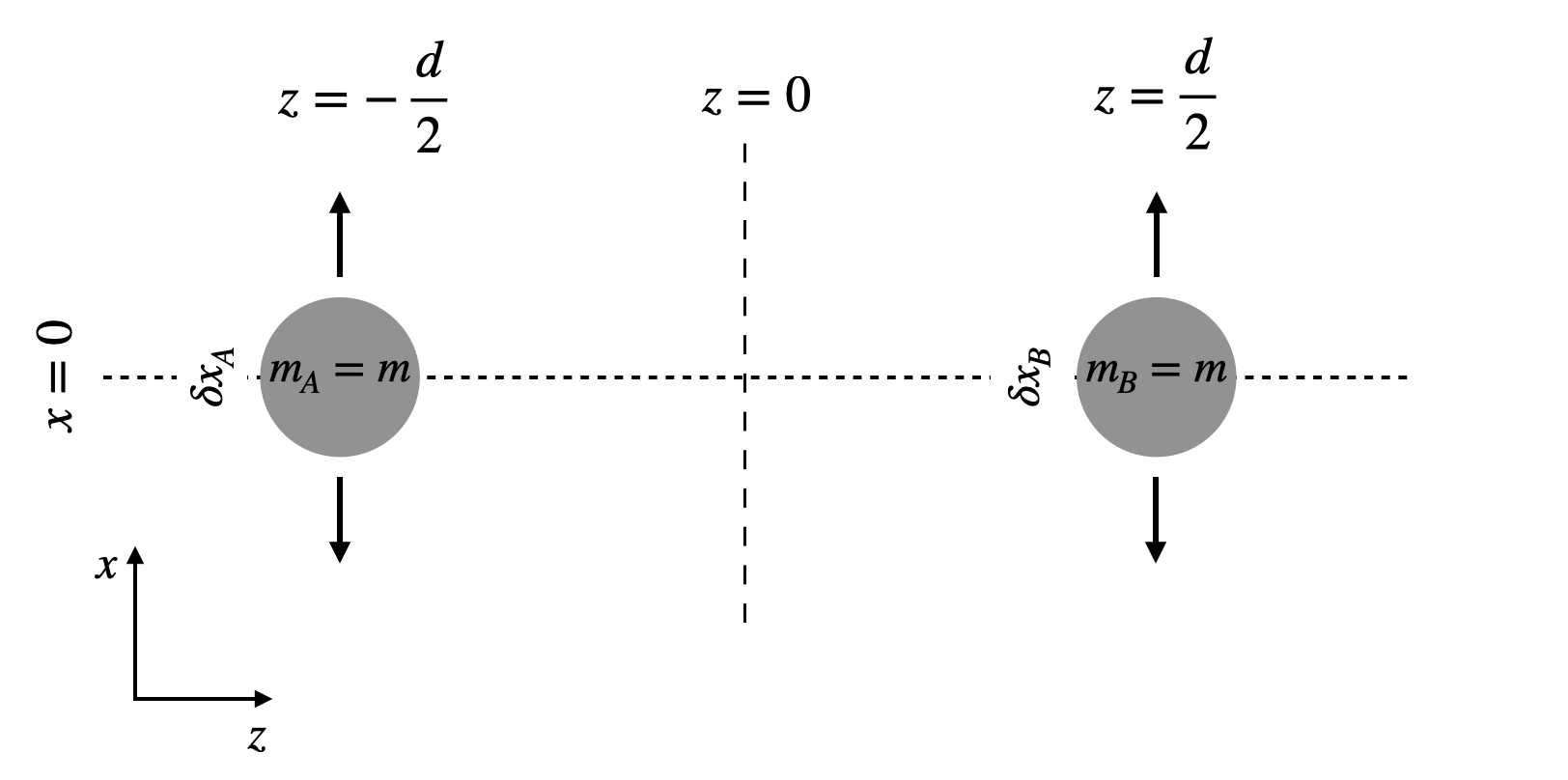}
   \caption{Illustration of the positions of particles A and B, separated by a distance $d$ along the $z$-direction and confined in harmonic potential traps. The two particles oscillate along the $x$-direction, with their respective displacements from the equilibrium positions denoted by $\delta\hat{x}_A$ and $\delta\hat{x}_B$. Both particles have the same mass $m$.}\label{fig:1}
\end{figure}

We consider two particles of mass $m$, labeled A and B, separated by a fixed distance $d$ along the $z$-direction. The two particles are confined in harmonic oscillator potentials, such that each particle oscillates along the $x$-direction, as illustrated in Fig. (\ref{fig:1}). The position operators of the two particles can be written as
\begin{eqnarray}
\hat{\boldsymbol{r}}_A=\left(\delta\hat{x}_A,\,0,\,-\frac{d}{2}\right)\,\,\,\,\,\,\,\,\,\,\,\,\,\,\,\,\,\,\,\,\,\,\,\,\,\,\,\,\,\,\,\,\,\,
\hat{\boldsymbol{r}}_B=\left(\delta\hat{x}_B,\,0,\,\frac{d}{2}\right),
\end{eqnarray}
where $\hat{\boldsymbol{r}}_A$ and $\hat{\boldsymbol{r}}_B$ denote the position operators of particles A and B, respectively, while $\delta\hat{x}_A$ and $\delta\hat{x}_B$ represent the displacements of the respective particles from their equilibrium positions due to their oscillatory motion.

Since both particles are confined in harmonic oscillator potentials along the $x$-direction, the Hamiltonian of the matter system can be written as
\begin{eqnarray}
\hat{H}_{\text{matter}}&=&\hat{H}_A+\hat{H}_B\nonumber\\
&=&\frac{\hat{p}^2_A}{2\,m}+\frac{m\,\omega^2_m}{2}\,\,\delta\hat{x}^2_A+\frac{\hat{p}^2_B}{2\,m}+\frac{m\,\omega^2_m}{2}\,\,\delta\hat{x}^2_B.
\end{eqnarray}
Here, $\omega_m$ is the oscillation frequency, assumed to be the same for both particles. The operators $\hat{p}_A$ and $\hat{p}_B$ are the momentum operators conjugate to $\delta \hat{x}_A$ and $\delta \hat{x}_B$, respectively. For convenience in the quantum description, the position and momentum operators can be expressed in terms of ladder operators (annihilation and creation operators) as
\begin{eqnarray}\label{eq:2.9}
\delta\,\hat{x}_A=\sqrt{\frac{1}{2\,m\,\omega_m}}\,\,\Big(\hat{a}+\hat{a}^{\dagger}\Big),&\,\,\,\,\,\,\,\,\,\,\,\,\,\,\,\,\,\,\,\,\,\,\,\,\,\,\,\,\,\,& \delta\,\hat{x}_B=\sqrt{\frac{1}{2\,m\,\omega_m}}\,\,\Big(\hat{b}+\hat{b}^{\dagger}\Big)\nonumber\\
\hat{p}_A=i\,\,\sqrt{\frac{m\,\omega_m}{2}}\,\,\Big(\hat{a}^{\dagger}-\hat{a}\Big),&\,\,\,\,\,\,\,\,\,\,\,\,\,\,\,\,\,\,\,\,\,\,\,\,\,\,\,\,\,\,\,\,& \hat{p}_B=i\,\,\sqrt{\frac{m\,\omega_m}{2}}\,\,\Big(\hat{b}^{\dagger}-\hat{b}\Big)
\end{eqnarray}
These ladder operators satisfy the canonical commutation relations $[\hat{a},\hat{a}^{\dagger}]=[\hat{b},\hat{b}^{\dagger}]=1$, while all other commutators vanish.

In this representation, the Hamiltonian of each particle takes the standard form of a quantum harmonic oscillator, and the energy eigenstates can be expressed in the occupation number basis. Consequently, the combined state of the matter system can be written in the tensor product basis $\ket{n_A}\otimes\ket{n_B}=\ket{n_A,n_B}$. These particles then interact with the quantized gravitational field described in the previous subsection. The form of the interaction Hamiltonian between the matter system and the gravitons will be derived and discussed in the following section.

\subsection{Graviton–Matter Interaction Hamiltonian}
The interaction Hamiltonian that couples the quantized gravitational field to the energy–momentum tensor of particles A and B can be derived from the general action of matter fields coupled to the metric,
\begin{eqnarray}
S_m=\int\,d^4x\,\sqrt{-g}\,T_{\mu\nu}\,\delta g^{\mu\nu}.
\end{eqnarray}
In the limit of small perturbations around a flat background metric, this action leads to the interaction Hamiltonian at linear order,
\begin{eqnarray}\label{eq:2.11}
\hat{H}_{\text{int}}&=&-\frac{1}{2}\,\int\,d^3x\,\,\hat{h}_{ij}(t,\boldsymbol{x})\,\,\hat{T}_{ij}(t,\,\boldsymbol{x})\nonumber\\
&=&-\frac{1}{M_{pl}}\,\sum_{\boldsymbol{k},s}\,\,\hat{h}^s_{\boldsymbol{k}}(t)\,\,\epsilon^s_{ij}(\boldsymbol{k})\,\,T_{ij}(t, -\boldsymbol{k})
\end{eqnarray}
where, in the second line, we have used the Fourier expansion of the gravitational field in equation (\ref{eq:2.3}) together with the Fourier transform of the energy–momentum tensor,
\begin{eqnarray}
\hat{T}_{ij}(t, \boldsymbol{k})=\frac{1}{\sqrt{V}}\,\int\,d^3 x\,\,e^{-i\boldsymbol{k}\cdot\boldsymbol{x}}\,\,\hat{T}_{ij}(t, \boldsymbol{x})
\end{eqnarray}

For the two-particle system oscillating along the $x$-direction with the geometry shown in Fig. (\ref{fig:1}), the dominant contribution to the coupling with the gravitational field in the configuration considered comes from the $T_{11}$ component of the energy-momentum tensor, which is associated with momentum fluctuations along the direction of particle oscillation. Therefore, the following analysis focuses on the interaction sector associated with this component, such that the effective gravitational interaction can be written as
\begin{eqnarray}\label{eq:2.13}
\hat{T}_{11}(t,-\boldsymbol{k})&=&\frac{1}{\sqrt{V}}\,\bigg(\frac{\hat{p}^{2}_A(t)}{E_A(t)}\,e^{-i\boldsymbol{k}\cdot\hat{\boldsymbol{r}}_A(t)}+\frac{\hat{p}^{2}_B(t)}{E_B(t)}\,e^{-i\boldsymbol{k}\cdot\hat{\boldsymbol{r}}_B(t)}\bigg).\nonumber\\
&=&\frac{1}{\sqrt{V}}\,\bigg(\frac{\hat{p}^{2}_A(t)}{E_A(t)}\,e^{-i\,\left(k_x\,\delta\hat{x}_A(t)-\,k_z\,\frac{d}{2}\right)}+\frac{\hat{p}^{2}_B(t)}{E_B(t)}\,e^{-i\left(k_x\,\delta\hat{x}_B(t)+\,k_z\,\frac{d}{2}\right)}\bigg).
\end{eqnarray}
It should be noted that the interaction form employed in this work is an effective form focused on the causal propagation contribution of the quantized gravitational field. Accordingly, the analysis in this work places greater emphasis on the role of gravitational-field propagation in the generation of quantum correlations between the two particles than on a complete treatment of the full tensorial structure of graviton interactions. Here, the operators $\hat{p}_{A/B}(t)$ and $\delta \hat{x}_{A/B}(t)$ denote the momentum and position operators in the interaction picture, evolving according to the free matter Hamiltonian,
\begin{eqnarray}\label{eq:2.14}
\hat{p}_{A/B}(t)=e^{i\hat{H}_{\text{matter}} t}\,\hat{p}_{A/B}\, e^{-i\hat{H}_{\text{matter}} t}, \,\,\,\,\,\,\,\,\,\,\,\,\,
\delta\hat{x}_{A/B}(t)= e^{i\hat{H}_{\text{matter}} t}\,\delta\hat{x}_{A/B}\,e^{-i\hat{H}_{\text{matter}} t}. 
\end{eqnarray}
Thus, the time dependence of the energy–momentum tensor is entirely determined by the free dynamics of the harmonic oscillators. The energy of each particle is expressed relativistically as $E_{A/B}=\sqrt{\hat{\boldsymbol{p}}^2_{A/B}+m^2}$. The exponential factors above represent phase operators depending on the particle positions in the interaction picture. The selection of the $\hat{T}_{11}$ component reflects the fact that only momentum fluctuations along the direction of oscillation contribute to the coupling with the gravitational modes in the configuration considered.

\section{Reduced Dynamics of the Matter System}\label{sec3}
Before examining how entanglement is generated from the coherent contribution of the quantized gravitational field with propagating modes, we first consider the reduced dynamics of the matter system after its interaction with these propagating modes of the gravitational field. This reduced dynamics can be derived using the Feynman–Vernon influence functional approach \cite{Breuer, Feynman, Feynman2}. In this approach, the gravitational field is treated as an environment for the matter system. By tracing out the gravitational degrees of freedom, the influence of the field as an environment on the matter system is expressed through the two-point correlation functions of the field, which are generally nonlocal in time. Within this framework, two generators naturally arise, namely $\hat{\mathcal K}^{(-)}$ and $\hat{\mathcal K}^{(+)}$, which originate from the commutator and anticommutator contributions of the gravitational field, respectively. The generator $\hat{\mathcal K}^{(-)}$ represents the coherent contribution associated with the causal response of the field and energy shifts, whereas $\hat{\mathcal K}^{(+)}$ describes the contribution from field fluctuations, which acts as noise and can lead to decoherence.

Let $\rho_{AB}(0)$ denote the initial density matrix of the matter system, and $\rho_G$ the initial state of the gravitational field (gravitons). The reduced dynamics of the matter system at time $t$ can be written as\begin{eqnarray}\label{eq:3.1}
\rho_{AB}(t)=\text{Tr}_{G}\bigg(\hat{U}(t,0)\,\,\Big(\rho_{AB}(0)\otimes\rho_G\Big)\bigg)
\end{eqnarray}
where $\mathrm{Tr}_{G}$ denotes the partial trace over the gravitational degrees of freedom. The time-evolution operator in the interaction picture is given by
\begin{eqnarray}
\hat{U}(t,0)=\mathcal{T}\,\,\text{exp}\bigg\{-i\,\int^t_0\,d\tau\,\hat{\mathcal{L}}(\tau)\bigg\},
\end{eqnarray}
with $\mathcal{T}$ denoting the time-ordering operator. Here, $\hat{\mathcal{L}}(t)$ is the Liouville superoperator defined through its action on a density matrix $\rho$ as
\begin{eqnarray}
\hat{\mathcal{L}}(t)\,\rho=\Big[\hat{H}_{\text{int}}(t),\,\rho\Big],
\end{eqnarray}
where $\hat{H}_{\text{int}}(t)$ is the interaction Hamiltonian given in equation (\ref{eq:2.11}).

To evaluate equation (\ref{eq:3.1}), the quantum state of the gravitational field is chosen to be the vacuum state, namely $\ket{0}_G$. Since the graviton vacuum is Gaussian, Wick’s theorem applies to the gravitational field operators appearing in the interaction Hamiltonian. As a consequence, all higher-order correlation functions can be reduced to products of two-point correlation functions. Within this framework, the reduced dynamics of the matter system can be expressed in an exponential form involving the two-point correlation function of the Liouville superoperator as follows
\begin{eqnarray}
\rho_{AB}(t)=\text{exp}\bigg\{\frac{1}{2}\,\int^t_0\,dt_1\,\int^t_0\,dt_2\,\,\,\text{Tr}_G\Big(\,\mathcal{T}\,\hat{\mathcal{L}}(t_1)\hat{\mathcal{L}}(t_2)\,\rho_G\Big)\bigg\}\,\rho_{AB}(0)
\end{eqnarray}
This exponential form is obtained by assuming that the graviton vacuum is Gaussian, so that Wick’s theorem applies to the gravitational field operators and all higher-order correlations can be reduced to products of two-point correlations. Under this condition, the cumulant expansion truncates at second order, yielding the exponential expression above. Then, by applying the time-ordering operator $\mathcal{T}$, the expression can be rewritten in the form of an ordered time integral as follows
\begin{eqnarray}\label{eq:3.5}
\rho_{AB}(t)=\text{exp}\bigg\{\int^t_0\,dt_1\,\int^{t_1}_0\,dt_2\,\,\,\text{Tr}_G\Big(\hat{\mathcal{L}}(t_1)\hat{\mathcal{L}}(t_2)\,\rho_G\Big)\bigg\}\,\rho_{AB}(0).
\end{eqnarray}
At this stage, equation (\ref{eq:3.5}) still retains the explicit dependence on the two times, $t_1$ and $t_2$, without applying any approximation to the dynamics of the matter system.

Next, attention is focused on the two-point correlation function of the Liouville superoperator, $\text{Tr}_G\big(\hat{\mathcal{L}}(t_1)\hat{\mathcal{L}}(t_2)\,\rho_G\big)$. By assuming that the gravitational field is in the vacuum state, its density matrix is given by $\rho_G=\ket{0}_G{}_G\bra{0}$. Using this density matrix and substituting the interaction Hamiltonian given in equation (\ref{eq:2.11}), the two-point correlation function can be written as
\begin{align}\label{eq:3.6}
&\text{Tr}_G\Big(\hat{\mathcal{L}}(t_1)\hat{\mathcal{L}}(t_2)\,\rho_G\Big)\nonumber\\
&=-\frac{1}{2M^2_{pl}}\,\sum_{\boldsymbol{k},s}\,\,\left|\epsilon^s_{11}(\boldsymbol{k})\right|^2\,\,\bigg({}_{G}\bra{0}\,\Big[\hat{h}^s_{-\boldsymbol{k}}(t_1),\,\hat{h}^s_{\boldsymbol{k}}(t_2)\Big]\ket{0}_G\Big(\hat{T}^{(L)}_{11}(t_1,\boldsymbol{k})-\hat{T}^{(R)}_{11}(t_1,\boldsymbol{k})\Big)\nonumber\\
&\,\,\,\,\,\,\,\times\Big(\hat{T}^{(L)}_{11}(t_2,-\boldsymbol{k})+\hat{T}^{(R)}_{11}(t_2,-\boldsymbol{k})\Big)+\,{}_{G}\bra{0}\,\Big\{\hat{h}^s_{-\boldsymbol{k}}(t_1),\,\hat{h}^s_{\boldsymbol{k}}(t_2)\Big\}\ket{0}_G\,\Big(\hat{T}^{(L)}_{11}(t_1,\boldsymbol{k})-\hat{T}^{(R)}_{11}(t_1,\boldsymbol{k})\Big)\nonumber\\
&\,\,\,\,\,\,\,\times\Big(\hat{T}^{(L)}_{11}(t_2,-\boldsymbol{k})-\hat{T}^{(R)}_{11}(t_2,-\boldsymbol{k})\Big)\bigg).
\end{align}
It can be seen that the two-point function of the Liouville superoperator naturally gives rise to the commutator and anticommutator contributions of the gravitational field. These two contributions represent the influence of the gravitational field as an environment on the matter system.

By substituting equation (\ref{eq:3.6}) into equation (\ref{eq:3.5}), the density matrix of the matter system at time $t$ becomes
\begin{eqnarray}\label{eq:3.7}
\rho_{AB}(t)=e^{-\hat{\mathcal{K}}^{(-)}(t)}\,\,e^{-\hat{\mathcal{K}}^{(+)}(t)}\,\rho_{AB}(0)
\end{eqnarray}
with
\begin{eqnarray}\label{eq:3.8}
\hat{\mathcal{K}}^{(-)}(t)&=&\frac{1}{2\,M^2_{pl}}\,\sum_{\boldsymbol{k},s}\,\,\left|\epsilon^s_{11}(\boldsymbol{k})\right|^2\,\,\int^t_0dt_1\,\int^{t_1}_0dt_2\,\,{}_{G}\bra{0}\,\Big[\hat{h}^s_{-\boldsymbol{k}}(t_1),\,\hat{h}^s_{\boldsymbol{k}}(t_2)\Big]\ket{0}_G\nonumber\\
&\,\,&\times\Big(\hat{T}^{(L)}_{11}(t_1,\boldsymbol{k})-\hat{T}^{(R)}_{11}(t_1,\boldsymbol{k})\Big)\Big(\hat{T}^{(L)}_{11}(t_2,-\boldsymbol{k})+\hat{T}^{(R)}_{11}(t_2,-\boldsymbol{k})\Big)
\end{eqnarray}
and
\begin{eqnarray}\label{eq:3.9}
\hat{\mathcal{K}}^{(+)}(t)&=&\frac{1}{2\,M^2_{pl}}\,\sum_{\boldsymbol{k},s}\,\,\left|\epsilon^s_{11}(\boldsymbol{k})\right|^2 \,\,\int^t_0dt_1\,\int^{t_1}_0dt_2\,\,{}_{G}\bra{0}\,\Big\{\hat{h}^s_{-\boldsymbol{k}}(t_1),\,\hat{h}^s_{\boldsymbol{k}}(t_2)\Big\}\ket{0}_G\,\nonumber\\
&\,\,&\times\Big(\hat{T}^{(L)}_{11}(t_1,\boldsymbol{k})-\hat{T}^{(R)}_{11}(t_1,\boldsymbol{k})\Big)\Big(\hat{T}^{(L)}_{11}(t_2,-\boldsymbol{k})-\hat{T}^{(R)}_{11}(t_2,-\boldsymbol{k})\Big)
\end{eqnarray}
Based on equation (\ref{eq:3.7}), the reduced dynamics of the matter system consists of two contributions arising from the correlations of the gravitational field, namely the commutator and anticommutator contributions. The commutator contribution, represented by $e^{-\hat{\mathcal{K}}^{(-)}(t)}$, is associated with the coherent part of the gravitational field and can give rise to energy shifts and nonlocal quantum correlations between particles A and B. Meanwhile, the anticommutator contribution, represented by $e^{-\hat{\mathcal{K}}^{(+)}(t)}$, is associated with field fluctuations and acts as a source of noise that can cause decoherence in the matter system. This separation of the two contributions is consistent with the analysis of Mitrakos et al. \cite{Mitrakos}. In the context of quantum gravity-induced entanglement, the field commutator is associated with the causal structure of the interaction, whereas the anticommutator is associated with quantum fluctuations of the field.

The expressions in equations (\ref{eq:3.8}) and (\ref{eq:3.9}) still retain the explicit dependence of the matter energy–momentum operator on the two times, $t_1$ and $t_2$. Subsequently, we assume that the dynamics of the matter system varies much more slowly than the dynamics of the gravitational field $(\omega_k \gg \omega_m)$. Under this condition, the variation of the matter energy–momentum operator over the correlation time of the gravitational field can be considered sufficiently small, allowing the approximation $\hat{T}_{11}(t_1,\boldsymbol{k})\approx\hat{T}_{11}(t_2,\boldsymbol{k})$ to be applied \footnote{This approximation is not intended to set $t_1-t_2=0$, but rather to assume that the variation of the matter operator over the time interval $t_1-t_2$ is sufficiently small.}. Based on this approximation, equations (\ref{eq:3.8}) and (\ref{eq:3.9}) can be simplified as
\begin{eqnarray}\label{eq:3.10}
\hat{\mathcal{K}}^{(-)}(t)&\approx&\frac{1}{2\,M^2_{pl}}\,\sum_{\boldsymbol{k},s}\,\, \left|\epsilon^s_{11}(\boldsymbol{k})\right|^2\,\,\int^t_0dt_1\,\int^{t_1}_0dt_2\,\,{}_{G}\bra{0}\,\Big[\hat{h}^s_{-\boldsymbol{k}}(t_1),\,\hat{h}^s_{\boldsymbol{k}}(t_2)\Big]\ket{0}_G\nonumber\\
&\,\,&\times\Big(\hat{T}^{(L)}_{11}(t_2,\boldsymbol{k})\hat{T}^{(L)}_{11}(t_2,-\boldsymbol{k})-\hat{T}^{(R)}_{11}(t_2,\boldsymbol{k})\hat{T}^{(R)}_{11}(t_2,-\boldsymbol{k})\Big)
\end{eqnarray}
and
\begin{eqnarray}\label{eq:3.11}
\hat{\mathcal{K}}^{(+)}(t)&\approx&\frac{1}{2\,M^2_{pl}}\,\sum_{\boldsymbol{k},s}\,\,\left|\epsilon^s_{11}(\boldsymbol{k})\right|^2 \,\,\int^t_0dt_1\,\int^{t_1}_0dt_2\,\,{}_{G}\bra{0}\,\Big\{\hat{h}^s_{-\boldsymbol{k}}(t_1),\,\hat{h}^s_{\boldsymbol{k}}(t_2)\Big\}\ket{0}_G\,\nonumber\\
&\,\,&\times\Big(\hat{T}^{(L)}_{11}(t_2,\boldsymbol{k})-\hat{T}^{(R)}_{11}(t_2,\boldsymbol{k})\Big)\Big(\hat{T}^{(L)}_{11}(t_2,-\boldsymbol{k})-\hat{T}^{(R)}_{11}(t_2,-\boldsymbol{k})\Big).
\end{eqnarray}
In this work, our aim is to analyze the causal structure underlying the generation of entanglement due to the interaction with the propagating modes of the gravitational field. As discussed above, this causal structure is encoded in the commutator contribution, or the coherent contribution of the gravitational field, represented by the operator $e^{-\hat{\mathcal{K}}^{(-)}(t)}$. Therefore, in the entanglement analysis presented in the following section, we specifically consider the contribution arising from the coherent part of the gravitational field and do not include the field-fluctuation contribution represented by $e^{-\hat{\mathcal{K}}^{(+)}(t)}$.

\section{Entanglement from the Coherent Contribution of the Quantized Gravitational Field}\label{sec4}
In this section, we investigate how the coherent contribution of the gravitational field generates entanglement between particles A and B. The matter-system dynamics associated with this contribution can be written in terms of the operator $e^{-\hat{\mathcal K}^{(-)}(t)}$, and the corresponding density matrix is given by
\begin{eqnarray}\label{eq:4.1}
\rho^{\text{coherent}}_{AB}(t)&=&e^{-\hat{\mathcal{K}}^{(-)}(t)}\,\rho_{AB}(0)\nonumber\\
&=&e^{-\hat{K}^{(-)}(t)}\,\rho_{AB}(0)\,e^{\hat{K}^{(-)}(t)}
\end{eqnarray}
In the second line, we have used the fact that the superoperator (\ref{eq:3.10}) has a commutator structure, namely $\hat{\mathcal{K}}^{(-)}(t)\,\rho = [\hat{K}^{(-)}(t),\,\rho]$. This allows the exponential action of the superoperator to be written in an operator form that sandwiches the density matrix between $e^{-\hat{K}^{(-)}(t)}$ and $e^{\hat{K}^{(-)}(t)}$, with
\begin{equation}\label{eq:4.2}
\hat{K}^{(-)}(t)=\frac{1}{2\,M^2_{pl}}\,\sum_{\boldsymbol{k},s}\,\,\left|\epsilon^s_{11}(\boldsymbol{k})\right|^2 \,\,\int^t_0dt_1\,\int^{t_1}_0dt_2\,\,{}_{G}\bra{0}\,\Big[\hat{h}^s_{-\boldsymbol{k}}(t_1),\,\hat{h}^s_{\boldsymbol{k}}(t_2)\Big]\ket{0}_G\,\,\hat{T}_{11}(t_2,\boldsymbol{k})\,\,\hat{T}_{11}(t_2,-\boldsymbol{k}),
\end{equation}
which can be interpreted as an effective operator associated with the superoperator $\hat{\mathcal{K}}^{(-)}(t)$. If the initial state of particles A and B is written in the general form $\ket{\psi_{AB}(0)}$, then the state of the system after interaction with gravitons is given by
\begin{eqnarray}\label{eq:4.3}
\ket{\psi^{\text{coherent}}_{AB}(t)}&=&e^{-\hat{K}^{(-)}(t)}\ket{\psi_{AB}(0)}\nonumber\\
&\approx&\Big(1-\hat{K}^{(-)}(t)\Big)\ket{\psi_{AB}(0)},
\end{eqnarray}
where the approximation in the second line is obtained in the weak-coupling regime, where the exponential operator can be expanded perturbatively to first order in the gravitational interaction.

Before discussing equation (\ref{eq:4.3}) further, we first determine the explicit form of the operator $\hat{K}^{(-)}(t)$. The first step is to calculate the commutator of the gravitational field $\hat{h}^s_{\boldsymbol{k}}(t)$ appearing in this operator. By substituting equation (\ref{eq:2.6}), we obtain
\begin{eqnarray}\label{eq:4.4}
{}_{G}\bra{0}\,\Big[\hat{h}^s_{-\boldsymbol{k}}(t_1),\,\hat{h}^s_{\boldsymbol{k}}(t_2)\Big]\ket{0}_G=-\frac{i}{k}\,\sin\Big(k\,\big(t_1-t_2\big)\Big).
\end{eqnarray}
Next, we consider the product of the energy-momentum tensors $\hat{T}_{11}(t_2,\boldsymbol{k})\,\hat{T}_{11}(t_2,-\boldsymbol{k})$. Here, we assume that particles A and B are in the nonrelativistic limit, namely $E_{A/B}\approx m$. Using equation (\ref{eq:2.14}), we obtain
\begin{eqnarray}\label{eq:4.5}
\hat{T}_{11}(t_2,\boldsymbol{k})\,\hat{T}_{11}(t_2,-\boldsymbol{k})&\approx&\frac{1}{V}\,\bigg(\frac{\hat{p}^2_A(t_2)}{m}\,e^{i\frac{\boldsymbol{k}\cdot\boldsymbol{d}}{2}}+\frac{\hat{p}^2_B(t_2)}{m}\,e^{-i\frac{\boldsymbol{k}\cdot\boldsymbol{d}}{2}}\bigg)\bigg(\frac{\hat{p}^2_A(t_2)}{m}\,e^{-i\frac{\boldsymbol{k}\cdot\boldsymbol{d}}{2}}+\frac{\hat{p}^2_B(t_2)}{m}\,e^{i\frac{\boldsymbol{k}\cdot\boldsymbol{d}}{2}}\bigg)\nonumber\\
&\approx&\frac{\hat{p}^2_A(t_2)\,\hat{p}^2_B(t_2)}{V\,m^2}\,\bigg(e^{i\,\left(\boldsymbol{k}\cdot\hat{\boldsymbol{r}}_A-\boldsymbol{k}\cdot\hat{\boldsymbol{r}}_B\right)}+e^{-i\,\left(\boldsymbol{k}\cdot\hat{\boldsymbol{r}}_A-\boldsymbol{k}\cdot\hat{\boldsymbol{r}}_B\right)}\bigg)
\end{eqnarray}
At this stage, the self-interaction terms are neglected, and only the cross terms are retained, since only these terms couple particles A and B and contribute to entanglement generation. Thus, for the propagating modes of the gravitational field, the lowest-order matter–matter interaction involves the momentum operators $\hat{p}_A$ and $\hat{p}_B$, with the spatial dependence expressed through the position operators $\hat{\boldsymbol{r}}_A$ and $\hat{\boldsymbol{r}}_B$.

By substituting equations (\ref{eq:4.4}) and (\ref{eq:4.5}) into equation (\ref{eq:4.2}), we obtain
\begin{align}\label{eq:4.6}
\hat{K}^{(-)}(t)=-\frac{i}{2\,m^2\,M^2_{pl}}\,\,&\int^t_0dt_1\,\int^{t_1}_0dt_2\,\,\hat{p}^2_A(t_2)\,\hat{p}^2_B(t_2)\,\,\nonumber\\
&\times\frac{1}{V} \,\,\sum_{\boldsymbol{k},s}\,\,\left|\epsilon^s_{11}(\boldsymbol{k})\right|^2 \,\frac{\sin\Big(k\,\big(t_1-t_2\big)\Big)}{k}\,\bigg(e^{i\,\left(\boldsymbol{k}\cdot\hat{\boldsymbol{r}}_A-\boldsymbol{k}\cdot\hat{\boldsymbol{r}}_B\right)}+e^{-i\,\left(\boldsymbol{k}\cdot\hat{\boldsymbol{r}}_A-\boldsymbol{k}\cdot\hat{\boldsymbol{r}}_B\right)}\bigg).\nonumber\\
\end{align}
Next, since our aim is to investigate the causal structure of entanglement generation arising from the coherent contribution of the propagating gravitational field, as well as to examine the role of the separation distance between the two particles in this process, we specifically restrict our analysis to gravitational-field modes propagating parallel to the separation axis of the two particles (along the $z$-direction) \footnote{It should be noted that this restriction is an additional assumption in our model. We do not assume that gravitational-field modes propagating in directions other than the $z$-direction make no contribution to the interaction. Rather, our analysis specifically focuses on modes propagating parallel to the separation axis of the two particles to investigate how the one-dimensional separation distance affects the causal structure of entanglement generation.}. Under this restriction, we can approximate $k \equiv |\boldsymbol{k}| \approx |k_z|$, while $\boldsymbol{k}\cdot\hat{\boldsymbol{r}}_A-\boldsymbol{k}\cdot\hat{\boldsymbol{r}}_B\approx k_z\,d$, and the polarization sum satisfies $\sum_s|\epsilon^s_{11}(\boldsymbol{k})|^2\approx 1/2$. We then take the continuum limit ($V\to\infty$), such that the sum over the wave vector $\boldsymbol{k}$ becomes an integral of the form

\begin{eqnarray}
\frac{1}{V}\,\sum_{\boldsymbol{k}}\,\,\,\,\,\,\,\,\,\to\,\,\,\,\,\,\,\,\,\frac{1}{A_{\perp}}\,\int\,\frac{dk_z}{2\pi},
\end{eqnarray}
where $A_{\perp}$ denotes the effective transverse area, defined as $A_{\perp}\equiv L_xL_y$. Since the propagating modes of the gravitational field are considered only along the $z$-direction, the integral involves only the $k_z$ component. Thus, the sum over the wave vector in equation (\ref{eq:4.6}) can be written as
\begin{align}\label{eq:4.8}
\frac{1}{V} \,\,\sum_{\boldsymbol{k},s}\,\,\left|\epsilon^s_{11}(\boldsymbol{k})\right|^2 &\,\,\frac{\sin\Big(k\,\big(t_1-t_2\big)\Big)}{k}\,\bigg(e^{i\,\left(\boldsymbol{k}\cdot\hat{\boldsymbol{r}}_A-\boldsymbol{k}\cdot\hat{\boldsymbol{r}}_B\right)}+e^{-i\,\left(\boldsymbol{k}\cdot\hat{\boldsymbol{r}}_A-\boldsymbol{k}\cdot\hat{\boldsymbol{r}}_B\right)}\bigg),\nonumber\\
&\approx\frac{1}{2\pi\,A_{\perp}}\,\int\,dk_z\,\,\frac{\sin\Big(|k_z|\,\big(t_1-t_2\big)\Big)}{|k_z|}\,\,\cos (k_z\,d)\nonumber\\
&=\frac{1}{2\,A_{\perp}}\,\,\Theta(t_1-t_2-d),
\end{align}
where $\Theta(x)$ denotes the Heaviside step function. The details of this integral calculation are presented in Appendix \ref{apendA}.

Based on equation (\ref{eq:4.8}), the operator $\hat{K}^{(-)}(t)$ can be written as
\begin{eqnarray}\label{eq:4.9}
\hat{K}^{(-)}(t)&=&-i\left(\frac{2\,\pi\,G}{A_{\perp}\,m^2}\right)\,\,\int^t_0dt_1\,\int^{t_1}_0dt_2\,\,\,\hat{p}^2_A(t_2)\,\hat{p}^2_B(t_2)\,\,\,\Theta(t_1-t_2-d)\nonumber\\
&=&-i\left(\frac{2\,\pi\,G}{A_{\perp}\,m^2}\right)\,\,\int^t_ddt_1\,\int^{t_1-d}_0dt_2\,\,\,\hat{p}^2_A(t_2)\,\hat{p}^2_B(t_2)
\end{eqnarray}
This equation shows that the coherent contribution of the gravitational field gives rise to an interaction between particles A and B involving the squared momentum operators, $\hat{p}_A^2\hat{p}_B^2$. This structure originates from the $\hat{T}_{11}$ component of the energy-momentum tensor in the nonrelativistic limit, after retaining only the cross terms that couple the two particles and performing the integral over the gravitational modes propagating along the $z$-direction. This interaction between particles A and B involving the squared momentum operators can generate an entangled state between the two particles. In addition, the presence of the Heaviside step function $\Theta(t_1-t_2-d)$ in the operator $\hat{K}^{(-)}(t)$ indicates that the contribution of this operator appears only when $t_1-t_2\geq d$. Consequently, the upper limit of the integral over $t_2$ becomes $t_1-d$, while the lower limit of the integral over $t_1$ becomes $d$. Physically, this indicates that the entanglement generated from the contribution of $\hat{K}^{(-)}(t)$ can only arise after a time interval of $d$, or $d/c$ for $c\neq1$, even though the interaction between the gravitational field and particles A and B has been present since $t=0$ \footnote{This result does not contradict the approximation used previously, since the propagation time $d/c$ is much shorter than the timescale of the system. For the parameters used in this work, namely $d=10^{-9}\,\mathrm{m}$ and $\omega_m=10^8\,\mathrm{Hz}$, we obtain $d/c\simeq3.33\times10^{-18}\,\mathrm{s}$, whereas the matter timescale is $1/\omega_m=10^{-8}\,\mathrm{s}$. Thus, the propagation time $d/c$ is approximately $3.33\times10^{-10}$ times the matter timescale.}. The presence of this time delay is consistent with previous studies \cite{Christodoulou, Bengyat}, which indicate a causal structure in the generation of entanglement due to interactions with propagating modes of the gravitational field.

By using the operator $\hat{K}^{(-)}(t)$ in equation (\ref{eq:4.9}), the quantum state of particles A and B after the interaction can be obtained as
\begin{eqnarray}\label{eq:4.10}
\ket{\psi^{\text{coherent}}_{AB}(t)}=\ket{\psi_{AB}(0)}+i\left(\frac{2\,\pi\,G}{A_{\perp}\,m^2}\right)\,\,\int^t_ddt_1\,\int^{t_1-d}_0dt_2\,\,\,\hat{p}^2_A(t_2)\,\hat{p}^2_B(t_2)\,\ket{\psi_{AB}(0)}.
\end{eqnarray}
At this stage, the initial state of particles A and B is still written in the general form $\ket{\psi_{AB}(0)}$, which in principle may already be an entangled state before the interaction begins. However, to examine the contribution of the gravitational field as a mechanism for entanglement generation, it is necessary to consider an initially separable state. Therefore, in the following section, we consider two specific cases of initially separable states, namely the ground state and the squeezed state of particles A and B. The resulting entanglement for these two cases is then examined and quantified separately.

\subsection{Coherent Entanglement from Initially Separated Ground States}\label{sec4.1}
In this section, the explicit form of equation (\ref{eq:4.10}) is derived by choosing the initial state of particles A and B as separable ground states, namely $\ket{\psi_{AB}(0)}=\ket{0_A,0_B}$. By substituting this initial state, the state of the system at time $t$ can be written as
\begin{eqnarray}\label{eq:4.11}
\ket{\psi^{\text{coherent}}_{AB}(t)}&=&\frac{1}{\sqrt{\mathcal{N}}}\,\,\Bigg[\bigg(1-i\left(\frac{\pi\,G\,\omega^2_m}{2\,A_{\perp}}\right)\big(t-d\big)^2\bigg)\,\ket{0_A,0_B}-\left(\frac{\pi\,G\,\omega^2_m}{2\,A_{\perp}}\right)\nonumber\\
&\,\,&\times \bigg(\frac{t-d}{\sqrt{2}\omega_m}-\frac{i\left(1-e^{2i\,\omega_m\,(t-d)}\right)}{2\sqrt{2}\,\omega^2_m}\bigg)\,\Big(\ket{2_A,0_B}+\ket{0_A,2_B}\Big)+\left(\frac{\pi\,G\,\omega^2_m}{2\,A_{\perp}}\right)\nonumber\\
&\,\,&\times \bigg(\frac{t-d}{2\,\omega_m}-\frac{i\left(1-e^{2i\,\omega_m\,(t-d)}\right)}{8\,\omega^2_m}\bigg)\,\ket{2_A,2_B}\Bigg],
\end{eqnarray}
where the normalization factor is given by
\begin{equation}
\mathcal{N}=1+\left(\frac{\pi\,G\,\omega^2_m}{2\,A_{\perp}}\right)^2\,\,\bigg[\left(\omega^2_m\,(t-d)^2+\frac{3}{4}\right)\,\frac{(t-d)^2}{\omega^2_m}+\frac{1-\cos\left(2\omega\,(t-d)\right)}{4\,\omega^4_m}+\frac{1-\cos\left(4\omega\,(t-d)\right)}{32\,\omega^4_m}\bigg].
\end{equation}
The results show that, at the lowest order of interaction, the gravitational contribution induces coupling between the ground state and the second excited state of the harmonic oscillators. In addition, the resulting state depends only on the oscillator frequency and does not explicitly depend on the particle mass. This mass independence arises as a direct consequence of the geometric configuration of the matter system and the structure of its interaction with the propagating gravitational field described in Section (\ref{sec2}). In the non-relativistic limit, the relevant energy–momentum component is $\hat{T}_{11}(t,\boldsymbol{k})$, which is dominated by the factor $\hat{p}_{A/B}^{\,2}/m$. Since the mass-dependent factor in $\hat{p}_{A/B}^{\,2}$ is proportional to m, the mass dependence cancels out.

To quantify the entanglement generated by the coherent contribution of the gravitational field when the two matter oscillators are initially in separate ground states, we employ the generalized concurrence, or I-concurrence. This quantity is used because the system under consideration consists of two harmonic oscillators with infinite-dimensional Hilbert spaces, for which the standard concurrence \cite{Wootters, Chuang} developed for two-qubit systems is not suitable. For a bipartite pure state, the generalized concurrence, or I-concurrence, is defined as
\begin{eqnarray}\label{eq:4.13}
\mathcal{C}\equiv \sqrt{2\Big(1-\text{Tr}\big[\rho^2_A(t)\big]\Big)},
\end{eqnarray}
Here, $\rho_A(t)$ denotes the reduced density matrix of particle A. Further discussion of the I-concurrence can be found in Ref. \cite{Rungta, Rungta2, Guo}. For simplicity, hereafter we refer to the generalized concurrence or I-concurrence simply as concurrence.

In the weak gravitational interaction regime, one obtains
\begin{eqnarray}\label{eq:4.14}
\mathcal{C}\approx\sqrt{2\left(1-\frac{1+4\left(\frac{\pi\,G\,\omega^2_m}{2\,A_{\perp}}\right)^2\left(\frac{(t-d)^2}{4\,\omega^2_m}+\frac{\cos\left(4\omega\,(t-d)\right)}{32\,\omega^4_m}\right)}{\mathcal{N}^2}\right)}
\end{eqnarray}
The details of the calculation are provided in Appendix \ref{apenB1}. Furthermore, by expanding the above expression to leading order and noting that the oscillatory terms remain bounded while the quadratic contribution in $\mathcal{N}$ grows with time, one obtains
\begin{eqnarray}\label{eq:4.15}
\mathcal{C}\sim\bigg(\frac{\pi\,G\,\omega^2_m}{\sqrt{2}\,A_{\perp}}\bigg)\,\,\big(t-d\big)^2
\end{eqnarray}
Based on this expression, it can be seen that the magnitude of the concurrence is directly proportional to the square of the oscillator frequency and inversely proportional to the transverse area $A_{\perp}$. In addition, the concurrence does not explicitly depend on the mass of the oscillating particles \footnote{It should be emphasized that this mass independence is not a general feature of all mechanisms of entanglement generation through interactions with the quantized gravitational field. The mass dependence may differ for different interaction sectors, as illustrated by different interaction contributions considered in Ref. \cite{Bose}.}. This mass independence of the concurrence is a consequence of the interaction structure through the $T_{11}$ component, as discussed above. This result is consistent with the study by Bose et al. \cite{Bose}, in which one of the cases considered also yields an entanglement measure that is independent of the mass of the oscillators. Within the leading-order approximation of Eq. (\ref{eq:4.15}), the concurrence increases with the duration of the interaction after the entangling contribution begins to develop. When $t=d$ in natural units, or $t=d/c$ when $c$ is restored explicitly, we obtain $\mathcal{C}=0$. This condition indicates that the concurrence begins to develop after the propagation time required to traverse the separation between the two oscillators.

However, the magnitude of the concurrence obtained remains very small for the interaction parameters considered. As an illustration, for an interaction time $t\sim1\,\mathrm{s}\approx1.52\times10^{24}\,\mathrm{GeV}^{-1}$, a separation distance $d=10^{-9}\,\mathrm{m}\approx5.07\times10^{6}\,\mathrm{GeV}^{-1}$, an oscillator frequency $\omega_m\sim10^8\,\mathrm{Hz}\approx6.58\times10^{-17}\,\mathrm{GeV}$, and an effective transverse area $A_{\perp}$ taken to be proportional to the square of the particle separation, namely $A_{\perp}\equiv d^2$, we obtain $\mathcal{C}\sim10^{-36}$. This very small value is consistent with the weakness of the gravitational interaction between particles A and B mediated by the propagating modes of the quantized gravitational field. Thus, the entanglement contribution arising from the coherent contribution of the gravitational field remains very small for the configuration considered here. This limitation motivates the need for a mechanism to enhance the entanglement contribution arising from the coherent part of the gravitational field. In the following section, we show that this contribution can be enhanced by appropriately choosing the initial quantum states of particles A and B, in particular, by preparing both particles in squeezed states.

\subsection{Coherent Entanglement from Initially Separated Squeezed States}\label{sec4.2}
In this section, the concurrence of the entangled state generated by the coherent contribution of the gravitational field is computed when the initial states of particles A and B are chosen as separable squeezed states. A squeezed state is a non-classical state of a harmonic oscillator characterized by the reduction of uncertainty in one quadrature variable, accompanied by an increase in uncertainty in its conjugate variable in accordance with the Heisenberg uncertainty principle \cite{Knight, Fox, Weinberg}. Mathematically, this state is obtained by applying the squeezing operator to the ground state. For each particle, the squeezed state is given by $\ket{\xi_A}=\hat{S}_A(\xi)\ket{0_A}$ and $\ket{\xi_B}=\hat{S}_B(\xi)\ket{0_B}$, with the squeezing operators
\begin{eqnarray}
\hat{S}_A(\xi)=\exp\bigg\{ \frac{1}{2}\,\big(\xi^*\,\hat{a}^2-\xi\,\hat{a}^{\dagger\,2}\big)\bigg\},\,\,\,\,\,\,\,\,\,\,\,\,\,\,\,\,\,\,\text{and}\,\,\,\,\,\,\,\,\,\,\,\,\,\,\,\,\hat{S}_B(\xi)=\exp\bigg\{ \frac{1}{2}\,\big(\xi^*\,\hat{b}^2-\xi\,\hat{b}^{\dagger\,2}\big)\bigg\},
\end{eqnarray}
where $\xi = r e^{i\theta}$, with $r$ as the squeezing parameter and $\theta$ as the phase. In this analysis, it is assumed that the squeezing parameter and phase are identical for both particles. By choosing the initial state $\ket{\psi_{AB}(0)}=\ket{\xi_A}\otimes\ket{\xi_B}$, the state of the system at time $t$ is given by
\begin{equation}\label{eq:4.17}
\ket{\psi^{\text{coherent}}_{AB}(t)}=\frac{1}{\sqrt{\mathcal{N}}}\left[\ket{\xi_A}\otimes\ket{\xi_B}+i\left(\frac{\pi\,G}{A_{\perp}\,m^2}\right)\,\,\int^t_ddt_1\,\int^{t_1-d}_0dt_2\,\,\,\hat{p}^2_A(t_2)\ket{\xi_A}\otimes\hat{p}^2_B(t_2)\,\ket{\xi_B}\right].
\end{equation}
with the normalization factor
\begin{eqnarray}
\mathcal{N}&=&1+\left(\frac{\pi\,G}{A_{\perp}\,m^2}\right)^2\,\int^t_ddt'_1\,\int^{t'_1-d}_0dt'_2\,\int^t_ddt_1\,\int^{t_1-d}_0dt_2\,\,\bra{\xi_A}\hat{p}^2_A(t'_2)\,\hat{p}^2_A(t_2)\ket{\xi_A}\,\nonumber\\
&\,\,&\otimes\bra{\xi_B}\hat{p}^2_B(t'_2)\hat{p}^2_B(t_2)\ket{\xi_B}.
\end{eqnarray}
This quantum state is none other than an entangled state.

Using the above state, the concurrence in the weak interaction limit is obtained as
\begin{eqnarray}\label{eq:4.19}
\mathcal{C}\approx\sqrt{2\left(1-\frac{1+\frac{1}{4}\,\left(\frac{\pi\,G\omega^2_m}{2\,A_{\perp}}\right)^2\,\big(t-d\big)^4\,\big(2\cosh^2(2r)\sinh^2(2r)+\cosh^4(2r)\big)}{\bigg(1+\frac{1}{4}\,\left(\frac{\pi\,G\omega^2_m}{2\,A_{\perp}}\right)^2\,\big(t-d\big)^4\,\big(\sinh^2(2r)+\cosh^2(2r)\big)^2\bigg)^2}\right)}.
\end{eqnarray}
The details of the calculation are provided in Appendix \ref{apenB2}. By expanding the above expression to leading order, one obtains
\begin{eqnarray}
\mathcal{C}\sim\bigg(\frac{\pi\,G\,\omega^2_m}{2\sqrt{2}\,A_{\perp}}\bigg)\,\,\big(t-d\big)^2\,e^{4r}
\end{eqnarray}
These results show that, compared with the ground-state case, the scaling structure of the concurrence remains similar but is modified by the exponential factor $e^{4r}/2$. The presence of this factor indicates that the magnitude of the entanglement arising from the coherent contribution of the gravitational field can be enhanced through the squeezing parameter. However, the resulting concurrence remains very small for the configuration considered here. As an illustration, for a squeezing parameter $r=1$, an interaction time $t\sim1\,\text{s}\approx1.52\times10^{24}\,\text{GeV}^{-1}$, a separation distance $d=10^{-9}\,\text{m}\approx5.07\times10^{6}\,\text{GeV}^{-1}$,an oscillator frequency $\omega_m\sim10^8\,\text{Hz}\approx6.58\times10^{-17}\,\text{GeV}$, and an effective transverse area defined as $A_\perp\equiv d^2$, we obtain $\mathcal{C}\sim10^{-34}$. Although the resulting concurrence remains very small, this result shows that the entanglement arising from the coherent contribution of the gravitational field can be enhanced through an appropriate choice of the initial quantum states. Therefore, increasing the squeezing parameter has the potential to further enhance the entanglement contribution arising from the coherent part of the gravitational field.

\section{Conclusion}\label{sec6}
In this paper, we have investigated how entanglement between two massive quantum particles is generated after they interact with propagating modes of the quantized gravitational field. The two particles are assumed to be trapped in harmonic oscillator potentials, to have the same mass and oscillation frequency, and to be separated by a certain distance. In addition, both particles are assumed to be in the non-relativistic limit. In this work, we specifically consider the coherent contribution of the gravitational field to investigate the causal structure and dynamics of entanglement generation. To analyze this, we employ the Feynman–Vernon influence functional approach.

Within the Feynman–Vernon influence functional approach, the gravitational field is treated as an environment interacting with the system, which in this case consists of two massive oscillating particles. The gravitational degrees of freedom are then traced out, yielding the quantum-state dynamics of the two massive particles. The influence of the gravitational field on the system dynamics is encoded in two generators associated with the commutator and anticommutator of the gravitational field, respectively. The generator associated with the commutator represents the coherent contribution of the gravitational field to the matter system and the causal structure of the interaction between the system and the environment. In this work, we specifically consider the entanglement generated by this coherent contribution of the gravitational field. Meanwhile, the generator associated with the anticommutator contributes to quantum noise, which can lead to decoherence. The contribution arising from the anticommutator-related generator is not analyzed further in this work. Therefore, the complete entanglement that includes both contributions is beyond the scope of the present work and may provide an interesting direction for future studies.

The entanglement arising from the coherent contribution of the gravitational field is investigated in the weak-interaction regime between the gravitational field and the matter system. In this regime, we find that entanglement does not arise instantaneously after the interaction begins. There is a time delay proportional to the distance between the two massive particles before entanglement begins to develop. This result shows that the entanglement generated from the coherent contribution of the gravitational field exhibits a causal structure.

However, in general, the entanglement arising from the coherent contribution of the gravitational field remains very small. As an illustration, when the initial quantum states of the two massive particles are chosen as separable ground states, the resulting entanglement, quantified by the concurrence, is of the order of $\sim 10^{-36}$. This concurrence can be enhanced when the initial quantum states of the two particles are prepared as separable squeezed states. In this case, the concurrence acquires an additional factor that depends exponentially on the squeezing parameter compared with the separable ground-state case. Thus, if squeezed states with suitable parameters can be realized, the resulting concurrence can be enhanced. Overall, this work investigates the generation of entanglement from the coherent contribution of the gravitational field within the framework of a quantized gravitational field with propagating modes, and demonstrates a causal structure characterized by a time delay in the generation of entanglement.

\section*{Acknowledgement}
F.P.Z. would like to thank PPMI FMIPA ITB and Kemendiktisaintek RI, the Ministry of Higher Education, Science, and Technology Republic of Indonesia for financial support.  A.T. would like to thank the members of Theoretical Physics Groups of Institut Teknologi Bandung for the hospitality.

\appendix

\section{Momentum Integral of the Graviton Commutator}\label{apendA}
In this section, we present in detail the evaluation of the integral in Eq. (\ref{eq:4.8}). To begin the discussion, we first consider the integral over $k_z$ in that equation. Since $\frac{1}{|k_z|}\,\sin\Big(|k_z|\,\big(t_1-t_2\big)\Big)\,\cos (k_z\,d)$ is an even function of $k_z$, the integration can be simplified as follows
\begin{align}\label{eq:A1}
\frac{1}{2\pi\,A_{\perp}}\,\int^{\infty}_{-\infty}\,dk_z\,\,&\frac{\sin\Big(|k_z|\,\big(t_1-t_2\big)\Big)}{|k_z|}\,\,\cos (k_z\,d)\nonumber\\
&=\frac{1}{\pi\,A_{\perp}}\,\int^{\infty}_0\,dk_z\,\,\frac{\sin\Big(k_z\,\big(t_1-t_2\big)\Big)}{k_z}\,\,\cos (k_z\,d)\nonumber\\
&=\frac{1}{2\pi\,A_{\perp}}\,\int^{\infty}_0\,dk_z\,\,\frac{1}{k_z}\,\,\left[\sin\Big((t_1-t_2+d)\,k_z\Big)+\sin\Big((t_1-t_2-d)\,k_z\Big)\right].
\end{align}
In the last step, we use the identity 
\begin{eqnarray}
\sin A \cos B=\frac{1}{2}\,\left(\sin(A+B)+\sin(A-B)\right).
\end{eqnarray}

Subsequently, the integral in Eq. (\ref{eq:A1}) can be evaluated using the Dirichlet integral, which can be expressed in terms of the Heaviside step function $\left(\Theta(x)\right)$, given by
\begin{eqnarray}
\int^{\infty}_0\,dx\,\,\frac{\sin(ax)}{x}=\frac{\pi}{2}\,\Big(2\,\Theta(x)-1\Big),
\end{eqnarray}
where $a$ is a real number. Using this relation, the integral in Eq. (\ref{eq:A1}) becomes
\begin{align}
\frac{1}{2\pi\,A_{\perp}}\,\int^{\infty}_{-\infty}\,dk_z\,\,&\frac{\sin\Big(|k_z|\,\big(t_1-t_2\big)\Big)}{|k_z|}\,\,\cos (k_z\,d)\nonumber\\
&=\frac{1}{4\,A_{\perp}}\,\left[\Big(2\,\Theta(t_1-t_2+d)-1\Big)+\Big(2\,\Theta(t_1-t_2-d)-1\Big)\right].
\end{align}
There are two Heaviside step functions in this expression, namely, $\Theta(t_1-t_2+d)$ and $\Theta(t_1-t_2-d)$. Since $t_1-t_2\gtrsim0$ and $d>0$, it follows that $t_1-t_2+d>0$, and therefore the Heaviside step function satisfies $\Theta(t_1-t_2+d)=1$. Thus, the integral in Eq. (\ref{eq:A1}) becomes
\begin{equation}
\frac{1}{2\pi\,A_{\perp}}\,\int^{\infty}_{-\infty}\,dk_z\,\,\frac{\sin\Big(|k_z|\,\big(t_1-t_2\big)\Big)}{|k_z|}\,\,\cos (k_z\,d)=\frac{1}{2\,A_{\perp}}\,\Theta(t_1-t_2-d).
\end{equation}
This result is precisely the result obtained in Eq. (\ref{eq:4.8}).

\section{Details of Concurrence Calculation in the Weak Interaction Limit}\label{apenB}
In this appendix, we present the detailed calculation of the concurrence for equations (\ref{eq:4.14}) and (\ref{eq:4.19}). The discussion is divided into two parts, with Appendix (\ref{apenB1}) containing the calculation for equation (\ref{eq:4.14}) and Appendix (\ref{apenB2}) containing the calculation for equation (\ref{eq:4.19}).

\subsection{Concurrence for Separable Ground State}\label{apenB1}
To compute the concurrence of the entangled state in equation (\ref{eq:4.11}), we first employ the weak interaction approximation, namely $\big(\frac{\pi\,G\,\omega^2_m}{2\,A_{\perp}}\big)\ll 1$. In this limit, the entangled state can be rewritten as
\begin{eqnarray}
\ket{\psi^{\text{coherent}}_{AB}(t)}\approx\frac{1}{\sqrt{\mathcal{N}}}\,\Bigg[\ket{0_A,0_B}+f(t)\,\Big(\ket{2_A,0_B}+\ket{0_A,2_B}\Big)-g(t)\,\ket{2_A,2_B}\Bigg],
\end{eqnarray}
with
\begin{eqnarray}
f(t)&=&\left(\frac{\pi\,G\,\omega^2_m}{2\,A_{\perp}}\right)\, \bigg(\frac{t-d}{2\,\omega_m}-\frac{i\left(1-e^{4i\,\omega_m\,(t-d)}\right)}{8\,\omega^2_m}\bigg),\label{eq:B.2}\\
g(t)&=&\left(\frac{\pi\,G\,\omega^2_m}{2\,A_{\perp}}\right)\, \bigg(\frac{t-d}{\sqrt{2}\,\omega_m}-\frac{i\left(1-e^{2i\,\omega_m\,(t-d)}\right)}{2\sqrt{2}\,\omega^2_m}\bigg).\label{eq:B.3}
\end{eqnarray}
To evaluate the concurrence, the reduced density matrix of one subsystem is required. Here we consider subsystem A, obtained by tracing out the degrees of freedom of subsystem B,
\begin{eqnarray}
\rho_A(t)&=&\text{Tr}_{B}\Big(\ket{\psi^{\text{coherent}}_{AB}(t)}\bra{\psi^{\text{coherent}}_{AB}(t)}\Big)\nonumber\\
&=&\frac{1}{\mathcal{N}}\,\,\Bigg[\Big(1+|f(t)|^2\Big)\,\ket{0_A}\bra{0_A}+\Big(f(t)-f^*(t)\,g(t)\Big)\,\ket{2_A}\bra{0_A}\nonumber\\
&\,\,&\,\,\,\,\,\,\,\,\,\,\,+\Big(f^*(t)-f(t)\,g^*(t)\Big)\,\ket{0_A}\bra{2_A}+\Big(|f(t)|^2+|g(t)|^2\Big)\,\ket{2_A}\bra{2_A}\Bigg].
\end{eqnarray}

From this reduced density matrix, one obtains
\begin{eqnarray}
\text{Tr}\Big[\rho^2_A(t)\Big]&=&\frac{1}{\mathcal{N}^2}\,\bigg[\Big(1+|f(t)|^2\Big)^2+\Big(|f(t)|^2+|g(t)|^2\Big)^2+2\,\Big|f(t)-g(t)\,f^*(t)\Big|^2\bigg]\nonumber\\
&\approx&\frac{1}{\mathcal{N}^2}\,\bigg[1+4\,|f(t)|^2\bigg].
\end{eqnarray}
In the second line, the weak interaction approximation is used, retaining terms up to second order in $\big(\frac{\pi\,G\,\omega^2_m}{2\,A_{\perp}}\big)$. Substituting equations (\ref{eq:B.2}) and (\ref{eq:B.3}), we obtain
\begin{eqnarray}
\text{Tr}\Big[\rho^2_A(t)\Big]=\frac{1}{\mathcal{N}^2}\,\left[1+4\,\left(\frac{\pi\,G\,\omega^2_m}{2\,A_{\perp}}\right)\, \left(\frac{(t-d)^2}{4\,\omega^2_m}+\frac{\cos \big(4\,\omega_m\,(t-d)\big)}{32\,\omega^4_m}\right)\right].
\end{eqnarray}
Substituting this result into the definition of concurrence in equation (\ref{eq:4.13}), we obtain
\begin{eqnarray}
\mathcal{C}&=& \sqrt{2\Big(1-\text{Tr}\big[\rho^2_A(t)\big]\Big)}\nonumber\\
&\approx&\sqrt{2\left(1-\frac{1+4\,\left(\frac{\pi\,G\,\omega^2_m}{2\,A_{\perp}}\right)\, \left(\frac{(t-d)^2}{4\,\omega^2_m}+\frac{\cos \big(4\,\omega_m\,(t-d)\big)}{32\,\omega^4_m}\right)}{\mathcal{N}^2}\right)}.
\end{eqnarray}
This expression is identical to the concurrence given in equation (\ref{eq:4.14}). Based on this result, an order-of-magnitude estimate of the concurrence can be obtained for the entangled state generated when the initial states of particles A and B are separable ground states.

\subsection{Concurrence for Separable Squeezed State}\label{apenB2}
In this section, we present the calculation of the concurrence for equation (\ref{eq:4.19}), which is obtained from the entangled state in equation (\ref{eq:4.17}). The first step is to compute the reduced density matrix of this state. As in the previous case, we consider subsystem A, which is obtained by tracing out the degrees of freedom of subsystem B,
\begin{eqnarray}
\rho_A(t)&=&\text{Tr}_{B}\Big(\ket{\psi^{\text{coherent}}_{AB}(t)}\bra{\psi^{\text{coherent}}_{AB}(t)}\Big)\nonumber\\
&=&\frac{1}{\mathcal{N}}\,\Bigg[\ket{\xi_A}\bra{\xi_A}+\left(\frac{\pi\,G}{A_{\perp}\,m^2}\right)^2\,\,\int^t_ddt_1\,\int^{t_1-d}_0dt_2\,\int^t_ddt'_1\,\int^{t'_1-d}_0dt'_2\,\,\hat{p}^2_A(t_2)\ket{\xi_A}\bra{\xi_A}\hat{p}^{\dagger\,\,2}_A(t'_2)\nonumber\\
&\,\,&\otimes\,\bra{\xi_B}\hat{p}^{2}_B(t_2)\,\hat{p}^{\dagger\,\,2}_B(t'_2)\ket{\xi_B}+i\,\left(\frac{\pi\,G}{A_{\perp}\,m^2}\right)\,\Bigg(\int^t_ddt_1\,\int^{t_1-d}_0dt_2\,\,\hat{p}^2_A(t_2)\ket{\xi_A}\bra{\xi_A}\nonumber\\
&\,\,&\otimes\,\bra{\xi_B}\hat{p}^{2}_B(t_2)\ket{\xi_B}\,-\int^t_ddt'_1\,\int^{t'_1-d}_0dt'_2\,\,\ket{\xi_A}\bra{\xi_A}\hat{p}^{\dagger\,\,2}_A(t'_2)\,\otimes\bra{\xi_B}\hat{p}^{\dagger\,\,2}_B(t'_2)\ket{\xi_B}\Bigg)\Bigg].
\end{eqnarray}
Using the weak interaction approximation, $\text{Tr}[\rho_A^2(t)]$ can be evaluated up to order $\big(\frac{\pi\,G}{A_{\perp}\,m^2}\big)^2$, yielding
\begin{eqnarray}
\text{Tr}\Big[\rho^2_A(t)\Big]&\approx&\frac{1}{\mathcal{N}^2}\,\,\Bigg[1-2\,\left(\frac{\pi\,G}{A_{\perp}\,m^2}\right)^2\,\bigg(\int^t_ddt_1\,\int^{t_1-d}_0dt_2\,\,\bra{\xi_A}\hat{p}^{2}_A(t_2)\ket{\xi_A}\,\otimes\,\bra{\xi_B}\hat{p}^{2}_B(t_2)\ket{\xi_B}\bigg)^2\nonumber\\
&\,\,&+4\,\left(\frac{\pi\,G}{A_{\perp}\,m^2}\right)^2\,\int^t_ddt_1\,\int^{t_1-d}_0dt_2\,\int^t_ddt'_1\,\int^{t'_2}_0dt'_2\,\,\bra{\xi_A}\hat{p}^{2}_A(t_2)\ket{\xi_A}\,\bra{\xi_A}\hat{p}^{2}_A(t'_2)\ket{\xi_A}\nonumber\\
&\,\,&\otimes\,\bra{\xi_B}\hat{p}^{2}_B(t'_2)\,\hat{p}^{2}_B(t_2)\ket{\xi_B}\Bigg].
\end{eqnarray}

To evaluate the above expression, we use the definition of the momentum operator in the interaction picture given in equations (\ref{eq:2.14}) and (\ref{eq:2.9}), as well as the relations between the squeezing operator and the ladder operators,
\begin{align}
\hat{S}^{\dagger}_A(\xi)\,\hat{a}^{\dagger}\,\hat{S}_A(\xi)&=\hat{a}^{\dagger}\,\cosh r-\hat{a}\,e^{-i\theta}\,\sinh r,\,\,\,\,\,\,\,\,\,\hat{S}^{\dagger}_A(\xi)\,\hat{a}\,\hat{S}_A(\xi)=\hat{a}\,\cosh r-\hat{a}^{\dagger}\,e^{i\theta}\,\sinh r\nonumber\\ \nonumber\\
\hat{S}^{\dagger}_B(\xi)\,\hat{b}^{\dagger}\,\hat{S}_B(\xi)&=\hat{b}^{\dagger}\,\cosh r-\hat{b}\,e^{-i\theta}\,\sinh r,\,\,\,\,\,\,\,\,\,\hat{S}^{\dagger}_B(\xi)\,\hat{b}\,\hat{S}_B(\xi)=\hat{b}\,\cosh r-\hat{b}^{\dagger}\,e^{i\theta}\,\sinh r\nonumber
\end{align}
Using these relations, the expectation values $\bra{\xi_{A/B}}\hat{p}^{2}_{A/B}(t_2)\ket{\xi_{A/B}}$ and $\bra{\xi_B}\hat{p}^{2}_B(t'_2)\,\hat{p}^{2}_B(t_2)\ket{\xi_B}$ can be evaluated. These expressions contain oscillatory terms in time. Since such terms remain bounded and do not grow with time, their contributions are neglected within the weak interaction approximation. Consequently, only the dominant terms that grow with time are retained. Under this approximation, one obtains
\begin{eqnarray}
\text{Tr}\Big[\rho^2_A(t)\Big]&\approx&\frac{1}{\mathcal{N}^2}\,\,\Bigg[1+\frac{1}{4}\,\left(\frac{\pi\,G\,\omega^2_m}{2\,A_{\perp}}\right)^2\,\big(t-d\big)^4\,\Big(2\cosh^2 (2r)\sinh^2 (2r)+\cosh^4(2r)\Big)\Bigg],
\end{eqnarray}
with the normalization factor, under the same approximation, given by
\begin{eqnarray}
\mathcal{N}\approx1+\frac{1}{4}\bigg(\frac{\pi\,G\,\omega^2_m}{4\,A_{\perp}}\bigg)^2\,\big(t-d\big)^4\,\Big(\sinh^2(2r)+\cosh^2(2r)\Big)^2.
\end{eqnarray}
Substituting this result into the definition of concurrence in equation (\ref{eq:4.13}), we obtain
\begin{eqnarray}
\mathcal{C}&=& \sqrt{2\Big(1-\text{Tr}\big[\rho^2_A(t)\big]\Big)}\nonumber\\
&\approx&\sqrt{2\left(1-\frac{1+\frac{1}{4}\,\left(\frac{\pi\,G\omega^2_m}{2\,A_{\perp}}\right)^2\,\big(t-d\big)^4\,\big(2\cosh^2(2r)\sinh^2(2r)+\cosh^4(2r)\big)}{\bigg(1+\frac{1}{4}\,\left(\frac{\pi\,G\omega^2_m}{2\,A_{\perp}}\right)^2\,\big(t-d\big)^4\,\big(\sinh^2(2r)+\cosh^2(2r)\big)^2\bigg)^2}\right)}.
\end{eqnarray}
This expression is identical to the concurrence in equation (\ref{eq:4.19}), and allows an order-of-magnitude estimate of the concurrence for the entangled state when the initial states of particles A and B are separable squeezed states.

\end{document}